\documentclass[preprint]{elsarticle}

\usepackage{lineno,hyperref}
\usepackage{psfrag,graphicx}
\usepackage{dcolumn}
\usepackage{amsmath,amssymb}
\usepackage{bm}
\usepackage{amsfonts,amssymb,amsmath}        
\usepackage{epstopdf}
\usepackage{enumerate}
\usepackage{float}

\modulolinenumbers[5]

\newcommand{\be}{\begin{equation}}
\newcommand{\ee}{\end{equation}}
\newcommand{\bq}{\begin{eqnarray}}
\newcommand{\eq}{\end{eqnarray}}

\newcommand{\ket}[1]{\left | \, #1 \right\rangle}

\makeatletter
\def\ps@pprintTitle{%
 \let\@oddhead\@empty
 \let\@evenhead\@empty
 \def\@oddfoot{}%
 \let\@evenfoot\@oddfoot}
\makeatother

\begin{document}

\begin{frontmatter}

\title{A simple decoder for topological codes}
\author{James R.~Wootton}
\address{Department of Physics, University of Basel, Klingelbergstrasse 82, CH-4056 Basel, Switzerland}
\date{}

\begin{abstract}

Here we study an efficient algorithm for decoding topological codes. It is a simple form of HDRG decoder, which could be straightforwardly generalized to complex decoding problems. Specific results are obtained for the planar code with both i.i.d. and spatially correlated errors. The method is shown to compare well with existing ones, despite its simplicity.

\end{abstract}

\end{frontmatter}


\section{Introduction}

Quantum error correcting codes are an essential part of proposals for quantum computation. Topological error correcting codes are currently one of the most well studied examples \cite{wootton_review}. No code is useful without a decoder, which takes the classical information produced by measurement of the code and uses it to determine how best to counteract the effects of noise. Multiple algorithms have been proposed to decode topological codes all with various advantages and disadvantages.

Here we present an algorithm whose main advantage is its simplicity, allowing straightforward generalization to a multitude of topological codes as well as more exotic decoding problems, such as non-Abelian decoding \cite{nonabelian}. It can be considered to be an example of an HDRG decoder, which have recently begun to attract attention \cite{broom,anwar,hdrg}, and is related to the `expanding diamonds' algorithm studied in \cite{dennis_thesis}.

To compare the method to existing ones we apply it to a specific choice of code and error model. The most widely studied topological codes are the surface codes \cite{dennis}. It is therefore these that we use for our benchmarking. We specifically use the planar code, in order to demonstrate the manner in which our algorithm deals with boundaries. Many decoding methods have been applied to this code \cite{hutter,fowler_new,max_like}, making it the best choice for comparing our algorithm.

The primary error model we consider is the standard one applied to the planar code. This has errors acting independently on each spin, and bit and phase flips occurring independently of each other. However, since realistic physical systems will likely experience errors that are correlated between spins \cite{novais,hutter_corr,fowler_corr}, we also consider an error model with nearest neighbour correlations between errors.

We also make use of a recent study of HDRG decoders \cite{hdrg}, of which our proposed method is an example. This work suggests an improved distance metric that can be used in such decoders in order to enhance performance for codes such as the planar code. We study the performance of our decoder both with and without this modification.

\section{The planar code}

\begin{figure}[t]
\begin{center}
{\includegraphics[width=6cm]{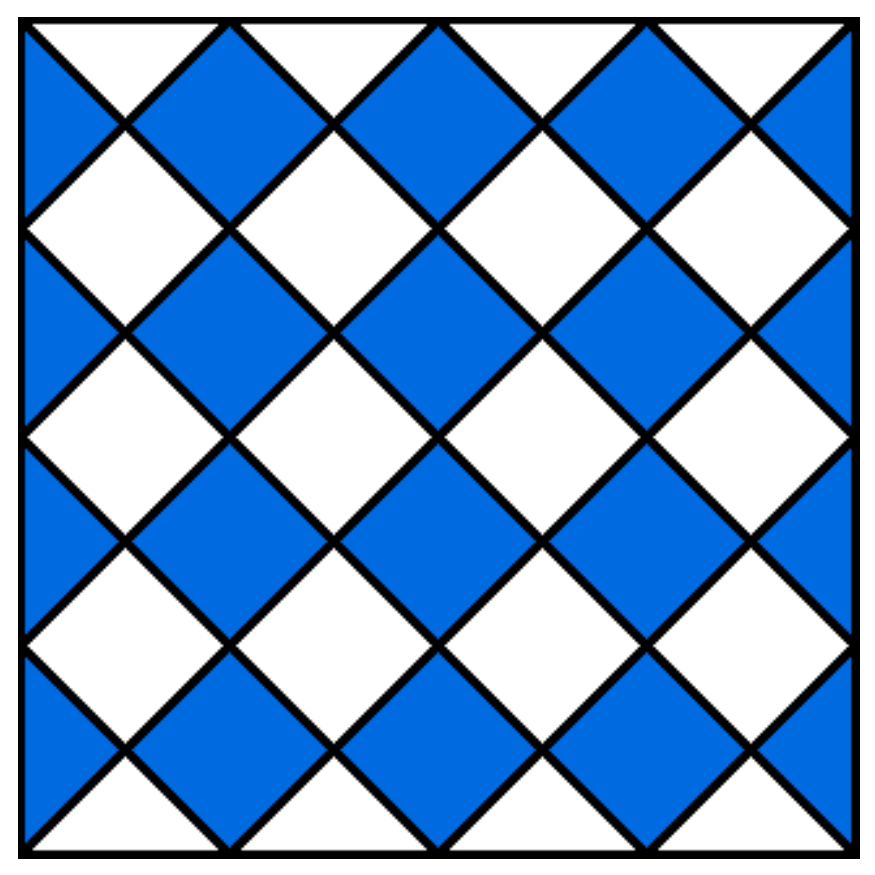}}
\caption{\label{fig1} The spin lattice of an $L \times L$ planar code. The $s$-plaquettes are shown in blue and the $p$-plaquettes are shown in white. A spin-$1/2$ particle is situated at each vertex. In this example the linear size is $L=5$. The linear size $L$ is also the distance of the code.}
\end{center}
\end{figure}

The planar code is the planar variant of the surface codes introduced by Kitaev \cite{double,dennis}. It is defined on a spin lattice, as in Fig. \ref{fig1}, with a spin-$1/2$ particle on each vertex. There are two types of plaquette, labelled $s$ and $p$, for which we define Hermitian operators as follows,
\be
A_s = \prod_{i \in s} \sigma^x_i, \,\,\, B_p = \prod_{i \in p} \sigma^z_i.
\ee
These operators mutually commute, and are the stabilizer operators of the code. Their eigenvalues can be interpreted in terms of anyonic occupations, with no anyon present for an eigenvalue of $+1$. Eigenvalues of $-1$ are interpreted as so-called flux anyons on $p$-plaquettes and charge anyons on $s$-plaquettes. The stabilizer space of the code therefore corresponds to the anyonic vacuum, and the anyon configuration is the syndrome of the code.

Edge operators may also be defined which commute with the stabilizers. These can be interpreted in terms of the anyon occupations of the edges, with flux anyons corresponding to a $-1$ eigenvalue of the left or right edge and charge anyons corresponding to a $-1$ eigenvalue of the top or bottom. The edge occupations define the state of the logical qubit stored in the code. The $X$ ($Z$) basis of the logical qubit can be defined such that the $\ket{+}$ ($\ket{0}$) state corresponds to vacuum on the top (left) edge and the $\ket{-}$ ($\ket{1}$) state corresponds to a flux (charge) anyon. Errors acting on the spins of the code will create and move anyons, and so disturb the stored information when they cause anyons to move off the edges. The job of decoding is to remove the anyons in a way that yields no net anyon moved off any edge. A logical error results when the decoding does this incorrectly.

The creation of pairs of plaquette anyons is achieved by bit flip errors. Pairs of vertex anyons are created by phase flips. The plaquette and vertex syndromes of the code can therefore be corrected independently. This will not give the best decoding in general, but it does represent a significant simplication. It is this approach that our decoder will take. Since the two syndrome types are equivalent, we restrict to plaquettes without loss of generality.

\section{The decoding algorithm}

The job of a decoding algorithm is to find a correction operator to neutralize the syndrome. This means that it must return the state of the code to one with trivial syndrome. In addition, the correction operator should also undo the effects of errors on the stored logical information. It is not possible to do this with unit probability. It can be done with high probability, with an exponential decay of failure rate with system size, but the success will depend greatly on the design of the decoder.

Ideally a decoder will perform a global optimization process to find the correction operator that is most likely to succeed, given the syndrome. However, such methods are never simple. An alternative is to start looking for patches of syndrome at small length scales that can be neutralized independently of the rest. Once all of the syndrome that can be dealt with at this scale is corrected, higher and higher scales can be considered until all of the syndrome is neutralized. Decoders that use this kind of iterative method are called HDRG decoders. See \cite{hdrg} for a general definition, and discussion of examples.

For the planar code, the syndrome is made up of anyons. These are created in pairs by strings of bit flip errors. Any pair of anyons can therefore be neutralized independently of the rest of the syndrome, with a string of bit flips serving as the required correction operator. Single anyons can also be neutralized by pairing them with an edge. Ideally the pairing used by correction is the same as that for creation, with the anyons annihilated in the pairs they were created in. It is not possible to do this perfectly, due to ambiguities in the syndrome information. However, we can instead aim to do it with high probability. As long as this results in the right parity (odd or even) for the number of anyons pairing with each edge, the correction of the logical information will succeed.

The decoder is designed to find pairs of anyons are likely to have been created by the same error string. The probability of this decays exponentially with their separation. As such, anyons that are mutual nearest neighbours will be a good candidate. The decoder therefore finds and pairs such mutual nearest neighbours. The method used to do this is as follows.

\begin{enumerate}

\item Loop through all plaquettes to find anyons.

\item For each anyon, search through all plaquettes at a Manhattan distance of $k$, using $k=1$ initially.

\item If either an anyon or the edge is found at this distance, pair them. Annihilate the anyons by performing bit flips on the connecting qubits. If there are multiple possibilities at this distance, pair with the first found.

\item If there are still anyons present, repeat the process for $k=k+1$.

\end{enumerate}

Once all the anyons are removed, the total pattern of bit flips used to remove them is considered. The correction procedure is a success if this belongs to the same equivalence class as the pattern of bit flips that occurred in error. Otherwise, the correction procedure results in a logical error.

An upper bound for the run time in the worst case scenario can be easily determined. The $k$th iteration requires $O(k)$ plaquettes to be searched per anyon. The number of anyons present during each iteration is at most $O(L^2)$. There will be no more than $L$ iterations since no anyon is more than this distance from the edge. The total complexity is then never more than $O(\sum_{k=1}^{L} k L^2) = O(L^4)$. This upper bound on the complexity is clearly polynomial with system size, and a moderately low ordered polynomial also. The algorithm therefore allows for fast and efficient decoding.

\section{Minimum requirements for logical errors} \label{min}

For any code and decoder, one important benchmark of performance is the minimum number of errors required to cause a logical error. We will use $\epsilon$ to denote the value of this number realized by an exhaustive decoder, and $\epsilon'$ to denote that for the decoder described above.

For the planar code, $\epsilon = L/2$ for even $L$ and $(L+1)/2$ for odd $L$. This is the minimum number of spin flips required to create a pair of anyons such that it takes less flips to pair them with opposite edges than with each other. Ignoring $O(1)$ corrections, as we will continue to do in the following, we can state this simply as $\epsilon = L/2$.

To find $\epsilon'$ we introduce the following terminology.
\begin{itemize}
\item A string is a collection of errors such that either an anyon or the edge exists at each endpoint.
\item A cluster is a collection of strings, and also of their corresponding endpoints.
\item The width, $w$, of a cluster is the maximum distance between a pair of its endpoints.
\item The distance, $d$, between non-overlapping clusters is the minimum distance between an endpoint from each. The clusters are said to overlap if there exists two endpoints $i$ and $j$ from one cluster and $k$ from the other such that $d_{ij}>d_{ik}$.
\item An isolated cluster is one whose width is smaller than the distance to any other cluster or to the edge.
\item A spanning cluster is one that contains both edges.
\item A level-$n+1$ cluster is a collection of level-$n$ clusters such that each is no further from another than its own width. A level-$0$ cluster is a single error.
\end{itemize}

Isolated clusters are defined such that no anyon within the cluster must look further than the width of the cluster to find a nearest neighbour before it is annihilated. Also no anyon within the cluster will be considered for pairing by any outside before it is annihilated. The algorithm will therefore annihilate the anyons in isolated clusters with each other (or with the edge if also included) without reference to any others.

Given a set of errors, consider all level-$0$ clusters. Any of these that are isolated clusters can all be ignored without affecting further analysis, because they have no effect on the other anyons.

By definition, all level-$0$ clusters that remain will be part of at least one level-$1$ cluster. Let us then consider all possible level-$1$ clusters. Any of these that are isolated will again be dealt with by the algorithm independently of the rest, and so can be ignored.

This procedure can repeated for higher level clusters until no errors remain. If no spanning cluster was ever considered, the algorithm will have certainly corrected the errors. The existence of a spanning cluster is therefore a necessary, but not sufficient, condition for a logical error. Determining the minimum number of errors required to create such a spanning cluster will then give a lower bound on $\epsilon'$.

For this it is clear that any errors that will end up being ignored before the cluster becomes spanning are superfluous. As such we require a set of errors in which none are ignored. All level-$0$ clusters (errors) must therefore be part of a level-$1$ cluster, all level-$1$ clusters part of a level-$2$ cluster, and so on.

Let us consider the uniform case, where each level-$n+1$ cluster contains $m$ level-$n$ clusters. In order to become spanning as soon as possible, it is clear that the best option is for all errors to be along a single line across the code. To make each level-$1$ cluster as long as possible there should be as many gaps between the $m$ errors as possible while maintaining the level-$1$ cluster. This would correspond to the $m$ errors being evenly spaced with a gap of one spin without an error between each, and so $m-1$ gaps in total. The width of each level-$1$ cluster will then be $w_1=2m-1$.

Similarly, to make each level-$2$ cluster as long as possible there should be $m-1$ gaps of length $w_1$ on which there are no errors. The width of each level-$2$ cluster will then be $w_2 = (2m-1)w_1 = (2m-1)^2$. Continuing in this way will lead to each level-$n$ cluster having a width of $w_n = (2m-1)^n$. Note also that each level-$n$ cluster will contain $m^n$ errors.

The minimum means to form a spanning cluster is to have a level-$n$ cluster where the left-most endpoint is closer to the left edge than to the right-most endpoint, and the right-most endpoint is closer to the right edge than to the left (or the same with left and right interchanged). The former ensures that the cluster is not isolated from the left edge, and the latter ensures that the both combined are not isolated from the right edge. The minimum width of a level-$n$ cluster required to satisfy this is $L/4$. This requires a cluster of level $n=\ln (L/4) / \ln (2m-1)$. The required number of errors is then $m^n = (L/4)^{\ln m / \ln (2m-1)}$. The exponent reaches its minimum value of $\beta = \log_3 2 \approx 0.63$ for $m=2$.

It is clear that the non-uniform case where different level-$n+1$ clusters can contain different numbers of level-$n$ clusters, will not allow a cluster to become spanning with fewer errors. As such $L^\beta$ is the minimum number of errors required for a spanning cluster, and hence forms a lower bound on $\epsilon'$.

For an upper bound on $\epsilon'$ we can find a set of errors that causes the algorithm to fail. One such example can be constructed similarly to the above. However, instead of the two level-$n$ clusters within a level-$n+1$ one being a distance $w_n$ apart, we instead put them $w_n-1$ apart. This will cause the algorithm to repeatedly pair the wrong anyons, and so cause a logical error.

Given this procedure it is easy to see that $w_{n} = 3 w_{n-1} - 1 = (3^n+1)/2$. This requires a cluster of level $n = \log_3 (L/2-1)$ for a logical error, and so $(L/2-1)^\beta$ errors. Combining this with the lower bound, we find that $\epsilon' = (cL)^\beta$ for $1/4 \leq c \leq 1/2$. Note that this is not linear with $L$, as we would ideally like, but its power law divergence would still be expected to provide good error suppression, as in the examples discussed in\cite{pryadko}. Note that the set of errors found here are related to the Cantor set \cite{dennis_thesis,landahl}.

\section{Numerical Results for Specific Error Models}

\subsection{Uncorrelated errors}

When benchmarking the planar code for a given decoder, it is typical to start with the case in which syndrome measurements are noiseless, and so all noise is due to errors acting on the spins of the code. Let us consider a general Pauli channel acting independently on each spin of the code, applying $\sigma_\alpha$ with probability $p_\alpha$ for $\alpha \in \{x,y,x\}$. Optimal decoding will fully take into account the exact values of these probabilities, and any correlations between the plaquette and vertex syndromes that they will cause. However, the decoder considered here takes the simpler route in which the two syndrome types are decoded independently. The plaquette decoding then depends only on the probability $p = p_x + p_y$ with which a bit flip, either $\sigma_x$ or $\sigma_z$, is applied. The vertex decoding similarly depends only on the probability $\tilde p = p_z + p_y$ for phase flips. Without loss of generality we therefore consider the error model in which bit and phase flips occur independently with the probabilities $p$ and $\tilde p$, respectively. We assume that $p> \tilde p$, so that we consider the worst of the two decoding problems.

\begin{figure}
\begin{center}
\includegraphics[width=\textwidth]{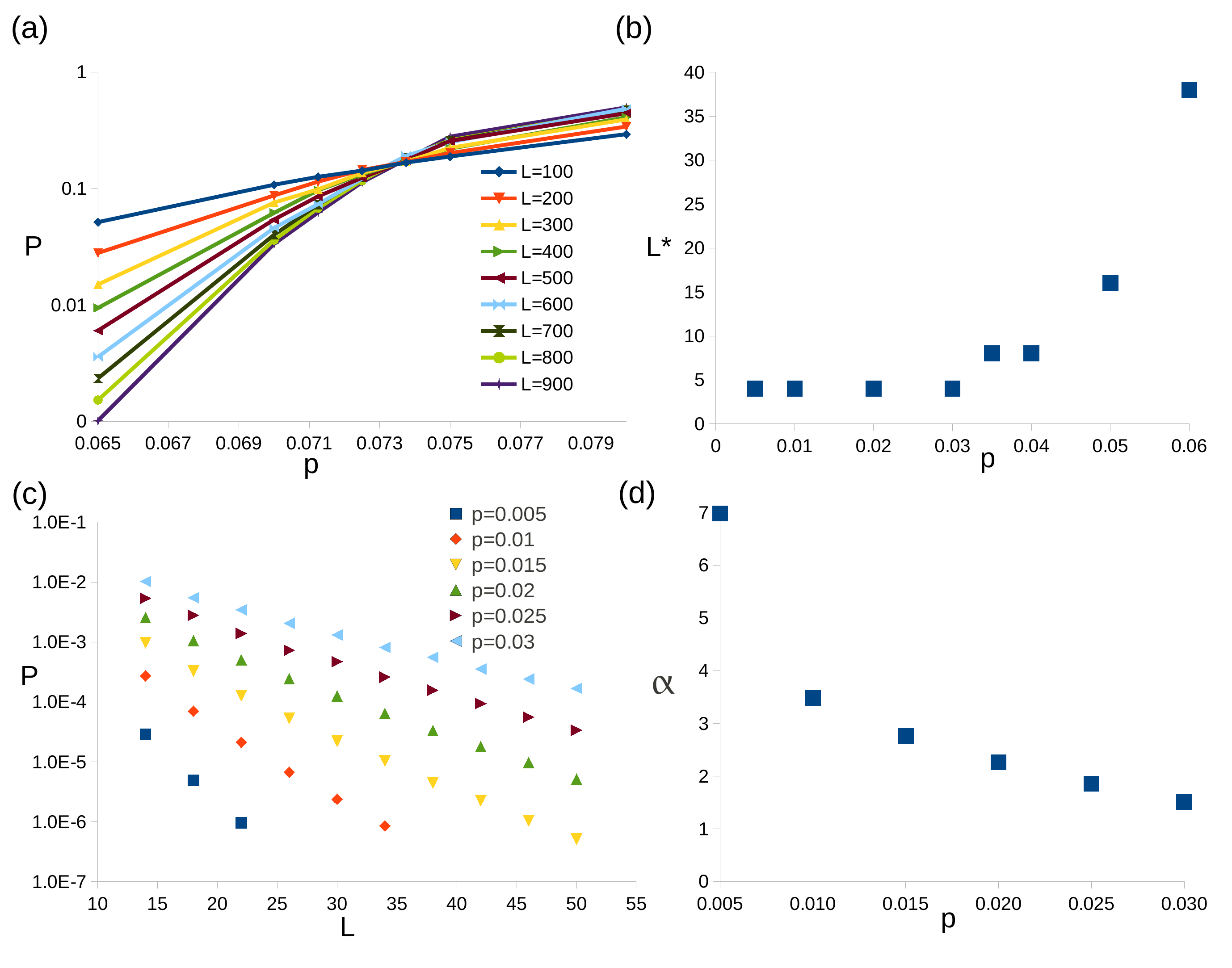}
\caption{\label{data}
Results for uncorrelated errors. (a) Graph of logical error rate, $P$, against bit-flip error rates, $p$, around threshold. (b) Graph of minimum linear system size required for $P<p$, $L^*$, against $p$. (c) Graph of $P$ against linear system size, $L$, for various $p$ well below threshold. (d) A graph of $\alpha(p)$ against $p$. }
\end{center}
\end{figure}

The decoder is benchmarked using the logical error rate, $P$, that it achieves for different bit flip error rates, $p$, and linear system sizes, $L$. This is done by randomly generating error configurations for the qubits of the code according to the noise model, applying the decoder to the resulting anyons, and then determining whether or not a logical error occurred for each sample. For each case we use the number of samples, $n$, required in order for $10^3$ logical errors to occur. The logical error rate is then $P=10^3/n$. The data was obtained to determine :

\begin{enumerate}[(a)]
\item The threshold error rate, $p_c$, for which $P$ decays with $L$ when $p<p_c$;
\item The minimum system size, $L^*$, required for each $p$ such that $P<p$ (and so the minimum system size required for a code stored qubit to be less noisy than one stored on a single physical spin);
\item The decay of the logical error rate for bit flip error rates well below threshold, to show that effective error correction occurs;
\item The values of $\alpha(p)$ for each $p$, when the above data is fitted to a function $P = 0(e^{-\alpha(p) (cL)^\beta})$.
\end{enumerate}
The results can all be found in Fig. \ref{data}. These suggest that the threshold is around $7.25-7.5 \, \%$. This is of the same order as the upper bound on the threshold of around $11\%$. It also compares well to the threshold of $7.8 \%$ obtained by the renormalization decoder of \cite{renorm}. It is larger than the $6.7 \%$ threshold obtained by the Bravyi-Haah algorithm \cite{broom}, but is less than the $8.4\%$ threshold of the modified version in \cite{anwar}. Note that these decoders deal with the toric code, which is identical to the planar code in the bulk but differs in its boundary conditions. Since the planar code boundary is known to lead to worse decoding behaviour \cite{fowler_toric}, it is remarkable that this decoder compares so well in its performance.

It is found that logical error rates of $P<p$ can be obtained using a small code of size $L=4$ for all $p \lesssim p_c/2$. The decoder therefore does not require a great deal of physical qubits in order to get good results. However, the required system size rises steeply for higher $p$. For $p \lesssim p_c/2$ we also find that the logical error rate decays very quickly as $L\rightarrow \infty$.

The decay below threshold will be of the form $P = 0(e^{-\alpha(p) (c L)^\beta})$. Here the scaling can be expected to be governed by the minimum requirements for a logical error. This means  $\beta = \log_3 2 \approx 0.63$ and, to be conservative, $c=1/4$.

To test this assumption, we can consider the scaling of $\log (-\log P)$ against $\log L$. This should be a linear relationship, with $\beta$ as the gradient. To minimize finite size effects, it is best to study low values of $p$. Specifically we consider $p = 0.1 \%$ and $p=0.05 \%$. Data was obtained for system sizes from $L=3$ to $L=11$, and linear interpolation was used to find the gradient. This gives $\beta = 0.637$ for $p = 0.1 \%$ and $\beta = 0.643$ for $p=0.05 \%$, with coefficient of determination $R^2=0.923$ and $R^2=0.919$ respectively. These results are consistent with the expected value of  $\beta \approx 0.63$.

To determine the values of the decay rate $\alpha(p)$, the data is fitted to a function of the form $P = 0(e^{-\alpha(p) (c L)^\beta})$ using $\beta = \log_3 2$. Note that this differs from the method used in \cite{bravyi}, but is morally the same.

\subsection{Spatially correlated errors}

We also consider errors that are not generated through a simple i.i.d. noise model, as above, but instead have spatial correlations. Such error models are arguably more realistic, and have recently attracted attention \cite{novais,hutter_corr,fowler_corr}. 

We consider a model for which an error occurs on each spin with probability $p'$. For each of these errors, an additional one will occur on a nearest neighbour with probability $q$. This error model is therefore equivalent to that above for $p'=p$ and $q=0$, but has nearest neighbour correlations between errors when $q>0$.

\begin{figure}
\begin{center}
\includegraphics[width=\textwidth]{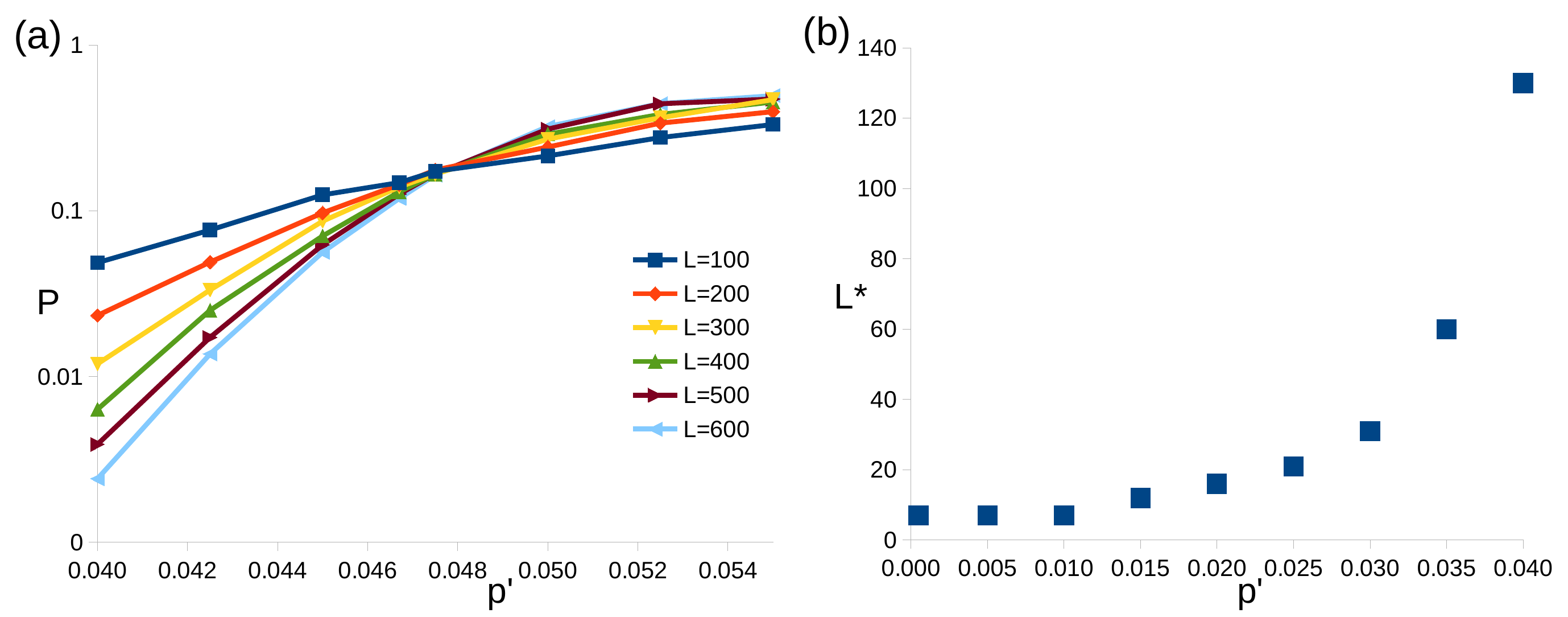}
\caption{\label{data2}
Results for correlated errors. (a) Graph of logical error rate, $P$, against bit-flip error rates, $p'$, around threshold. (b) Graph of minimum linear system size required for $P<p'$, $L^*$, against $p'$. }
\end{center}
\end{figure}

Using the same method of numerical benchmarking described above we study the threshold, $p_c'$, and the minimum system size required for error correction, $L^*(p)$, for the case of $q=0.5$. The results are shown in Fig. \ref{data2}.

To compare the results with those for the uncorrelated case, let us consider the expected fraction of spins to have suffered an error in both cases. For the uncorrelated case this is simply $p$, whereas in this case it will be approximately $(1+q)p'$. We may therefore expect that the behaviour of the decoder for given values of $p'$ and $q$ in this case will correspond to the behaviour for $p=(1+q)p'$ in the uncorrelated case. This gives us an estimate of $p_c' \approx p_c/1.5 = 4.8 \%$ for $q=0.5$. 

The threshold is found numerically to be around $4.75-5 \, \%$, and so agrees well with the above expectations. However the logical errors rates found in this case with a given $p'$ fall well below those for $p=(1+q)p'$ in the uncorrelated case. For example, uncorrelated $p = 6\%$ corresponds to correlated $p'=4\%$, however for $L=600$ we find $P = 1.6 \times 10^{-4}$ in the former case and $P = 2.4 \times 10^{-3}$ in the latter. The correlated case therefore gives worse logical error rates by an order of magnitude. Also the minimum system size for error correction to be evident is $L^*=38$ in the former case but $L^*=130$ in the latter. This decrease in performance is likely due to the fact that the correlations increase the typical length of error strings. This effectively decreases the distance of the code.

For the minimum required system sizes, $L^*$, required for $P<p'$, it is found that at least $L=7$ is required for error rates as low as $p'=5 \times 10^{-4}$. However, this small size remains sufficient until $p'=1 \%$, after which the required system size rises steeply as $p' \rightarrow p_c$.

\section{Modified distance metric}

The algorithm presented here is an example of an HDRG decoder. Some of the properties of this class of decoding methods were presented in \cite{hdrg}. This includes a proposed improvement for these decoders, aimed at increasing the minimum number of physical errors required to cause logical error.

The proposed improvement is a redefinition of the distances used by the decoder. The purpose of the distances is to reflect how likely two elements of the syndrome are to be endpoints of the same error string. The most obvious choice is therefore the number of errors required to correct them, which is the Manhattan distance for the planar code defined on a square lattice. However, this does not always lead to the best results.

In some iteration of the process, suppose the decoder chooses to annihilate two anyons $b$ and $c$, separated by a Manhattan distance $D_{bc}$. The decoder then assumes that a chain of $D_{bc}$ errors has occurred between them. At some later iteration, the decoder will need to determine whether another pair, $a$ and $d$, should be annihilated. For this it will need to know the number of errors an error chain between them would require. Ordinarily, this would be the Manhattan distance $D_{ad}$. However, consider a path between $a$ and $d$ that intersects $c$ and $b$. An error chain that goes from $a$ to $b$, and then from $c$ to $d$, would therefore also be consistent with the original syndrome. This would require $D_{ab}+D_{cd}$ errors, which may be less that $D_{ad}$ in general. The pairing of $a$ and $d$ would then be more likely than the decoder would assume due to the possibility of this shortcut for the error chain.

The improvement suggested in \cite{hdrg}, when adapted for the specific case of the planar code, is to perform an update of the distances between pairs of anyons each time an annihilation occurs. Initially the distance, $D(a,b)$ between all pairs of anyons will be set to be the Manhattan distance. Upon the annihilation of a pair, $c$ and $d$, the distances are updated according to
\be
D_{a,b} \rightarrow \min \left( D_{a,b} \, , \, D_{a,c} + D_{d,b} \, , \, D_{a,d} + D_{c,b} \right).
\ee
This will allow the syndrome information concerning pairs annihilated early in the process to still be used later in the process, even though they have been removed from the syndrome. As such, the decoder can make better decisions concerning which anyons to pair at later stages.

\begin{figure}
\begin{center}
\includegraphics[width=\textwidth]{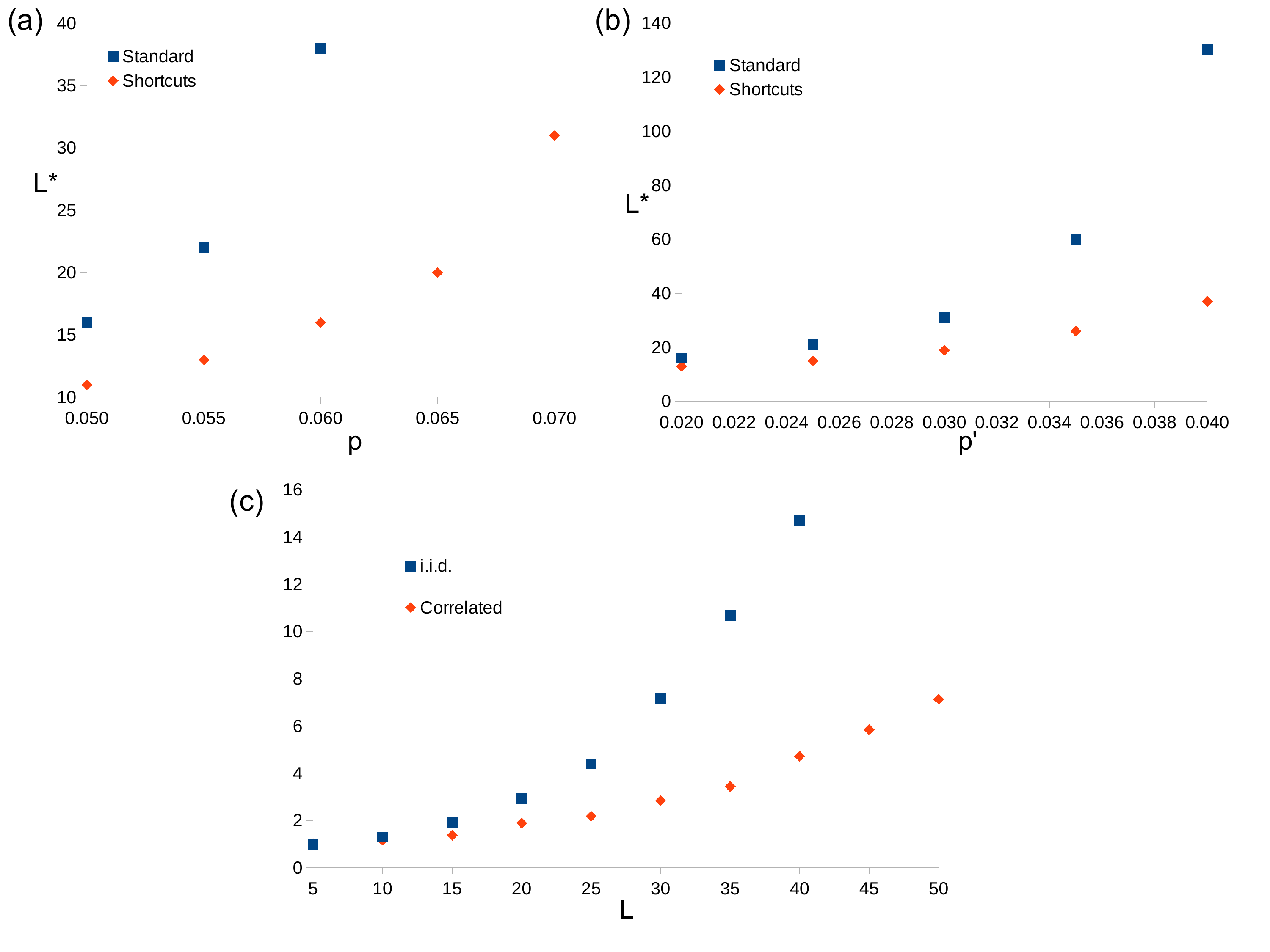}
\caption{\label{data3}
Comparison of the results of the method using shortcuts with the standard method. (a) Graphs of $L^*$ against $p$ for uncorrelated errors using both methods. (b) Graphs of $L^*$ against $p'$ for correlated errors using both methods. (b) Graph of $P(standard)/P(shortcuts)$ against $L$ for both uncorrelated and uncorrelated errors. For the former, the data is for $p=3.5 \%$. For the latter it is $p=2.5 \%$.}
\end{center}
\end{figure}

In Fig. \ref{data3} data is presented to compare the results of decoder when these redefined distances are used to that when they are not. These are referred to as the `shortcut' and `standard' methods, respectively. Due to the way the shortcut method is implemented, it is more difficult to obtain results for large system sizes. As such, a direct comparison of thresholds is not made. Instead the system size $L^*$ is compared for both the correlated and uncorrelated error models. It is found that the shortcuts offer no significant advantages for low error rates, for which the corresponding $L^*$ is small. However, for larger error rates the growth of $L^*$ with $p$ and $p'$ is found to be much slower when the shortcuts are used, and so the same degree of error suppression can be achieved with much smaller systems.

The ratio of logical error rates is also compared. For the uncorrelated error model the ratio $P(standard)/P(shortcuts)$ is found for $p=3.5 \% \approx p_c/2$ and multiple system sizes. This ratio quantifies how much more likely the standard method is to fail than the one with shortcuts. The results suggest that this increases exponentially with $L$, or a power thereof. This suggests that the value of $\beta$ for the shortcuts is higher than that for the standard method, as is expected \cite{hdrg}. The ratio is similarly found for the correlated error model and $p'=2.5 \% \approx p'_c/2$, with similar results.

\section{Conclusions and Outlook}

The standard version of the decoder presented here is very much the `vanilla' version. There are many changes that one could imagine making in order to improve the performance of the method, beyond the known example of shortcuts. For example, rather than pairing with the first anyon found in each case, one can compare all anyons found at the same distance and make an informed choice on which to pair with. Another possibility is the Markov chain Monte Carlo techniques enhancements of \cite{hutter}. All such modifications will depend on the code and error model considered. This paper only aims to demonstrate the core method, so such application specific modifications were not studied here. 

Despite its simplicity, the method is shown to behave very well as a decoder. It achieves respectable benchmarks for the standard testing ground of the planar code with independent bit and phase flip noise: the threshold is similar to other decoders and the theoretical maximum; error correction is evident for small system sizes; and the logical error rate is found to decay strongly.  Good results were also found for the case of spatially correlated errors. The decoder can therefore be used to gain good results for topological codes as well as more exotic decoding problems, before more complex tailor-made methods are developed.


\section{Acknowledgments}

The author would like to thank Benjamin Brown for discussions, Andrew Landahl for pointing out the prior existence of a variant of this decoder, and the Swiss NF, NCCR Nano and NCCR QSIT for support.




\begin{thebibliography}{999} 

\bibitem{wootton_review} Wootton, J. R. Quantum memories and error correction. \emph{J. Mod. Opt.} \textbf{2012}, \emph{20}, 1717.

\bibitem{nonabelian} Wootton, J. R.; Burri, J.; Iblisdir, S.; Loss, D. Error Correction for Non-Abelian Topological Quantum Computation. \emph{Phys. Rev. X} \textbf{2014}, \emph{4}, 011051.

\bibitem{broom} Bravyi, S.; Haah, J. Quantum Self-Correction in the 3D Cubic Code Model. \emph{Phys. Rev. Lett.} \textbf{2013}, \emph{111}, 200501.

\bibitem{anwar} Anwar, H.; et al. Fast decoders for qudit topological codes.		\emph{New J. of Phys.} \textbf{2014}, \emph{16}, 063038.

\bibitem{hdrg} Hutter, A.; Loss, D.; Wootton, J. R. Improved HDRG decoders for qudit and non-Abelian quantum error correction. \emph{arXiv:1410.4478}		\textbf{2014}.

\bibitem{dennis_thesis} Dennis, E. Purifying Quantum States: Quantum and Classical Algorithms	. \emph{arXiv:quant-ph/0503169} \textbf{2005}.

\bibitem{dennis} Dennis, E.; Kitaev, A.; Landahl, A.; Preskill, J. Topological quantum memory. \emph{J. Math. Phys.} \textbf{2002}, \emph{43}, 4452.

\bibitem{hutter} Hutter, A; Wootton J. R.; and Loss, D. Efficient Markov chain Monte Carlo algorithm for the surface code. \emph{Phys. Rev. A}	\textbf{2014}, \emph{89}, 022326.
\bibitem{fowler_new} Fowler, A. G.  Optimal complexity correction of correlated errors in the surface code. \emph{arXiv:1310.0863} \textbf{2013}.
\bibitem{max_like} Bravyi, S.; Suchara, M.; Vargo, A. Efficient algorithms for maximum likelihood decoding in the surface code.	\emph{Phys. Rev. A} \textbf{2014}, \emph{90}, 032326.

\bibitem{novais} Novais E.; Mucciolo, E. R. Surface Code Threshold in the Presence of Correlated Errors. \emph{Phys. Rev. Lett.} \textbf{2013}, \emph{110}, 10502.

\bibitem{hutter_corr} Hutter, A.; Loss, D. Breakdown of surface-code error correction due to coupling to a bosonic bath. \emph{Phys. Rev. A} \textbf{2014}, \emph{89}, 042334.

\bibitem{fowler_corr} Fowler, A. G.; Martinis, J. M. Quantifying the effects of local many-qubit errors and nonlocal two-qubit errors on the surface code. \emph{Phys. Rev. A} \textbf{2014}, \emph{89}, 03216.

\bibitem{double} Kitaev, A. Fault-tolerant quantum computation by anyons. \emph{Annals Phys.}	\textbf{2003}, \emph{303}, 2.

\bibitem{pryadko} Kovalev, A. A.; Pryadko, L. P. Fault tolerance of quantum low-density parity check codes with sublinear distance scaling. \emph{Phys. Rev. A} \textbf{2013}, \emph{87}, 020304(R).

\bibitem{renorm} Duclos-Cianci, G.; Poulin, D. Fast Decoders for Topological Quantum Codes. \emph{Phys. Rev. Lett.} \textbf{2010}, \emph{104}, 050504.

\bibitem{fowler_toric} Fowler, A. G. Accurate simulations of planar topological codes cannot use cyclic boundaries. \emph{Phys. Rev. A} \textbf{2013}, \emph{87}, 062320.

\bibitem{bravyi} Bravyi, S.; Vargo, A. Simulation of rare events in quantum error correction. \emph{Phys. Rev. A} \textbf{2013}, \emph{88}, 062308.

\bibitem{landahl} Landahl, A. Private communication. \textbf{2013}.



\end{thebibliography}
\end{document}